\documentclass[fleqn,usenatbib]{mnras}

\usepackage{newtxtext,newtxmath}

\usepackage[T1]{fontenc}
\usepackage{ae,aecompl}

\usepackage{graphicx}   
\usepackage{amsmath}    

\usepackage{color}
\usepackage{placeins}

\newcommand\be{\begin{equation}}
\newcommand\en{\end{equation}}
\newcommand\msun{M_{\odot}}
\newcommand\rsun{R_{\odot}}

\newcommand\mdot{\dot{M}}
\newcommand\msunyr{M_{\odot}\, {\rm yr^{-1}}}

\newcommand\gmcmtwo{\rm g\, cm^{-2}}
\newcommand\kms{\rm km s^{-1}}

\title[Disc outbursts ] {Magnetically-activated accretion outbursts of pre-main sequence discs}
  \author[Cleaver, Hartmann \& Bae]
      {Jacob Cleaver$^{1}$, Lee Hartmann$^{1}$, Jaehan Bae$^{2}$ \thanks{E-mail:
          lhartm@umich.edu} 
\\
$^{1}$Department of Astronomy, University of Michigan,  
	1085 S. University Ave, Ann Arbor, MI 48105, USA \\
$^{2}$ Department of Astronomy, University of Florida,
316 Bryant Space Science Building, Gainesville, FL 32611, USA \\
}

\date{Accepted XXX. Received YYY; in original form ZZZ}

\pubyear{2017}

\begin{document}
\label{firstpage}
\pagerange{\pageref{firstpage}--\pageref{lastpage}} \maketitle

\begin{abstract}
We investigate whether
triggering of the magnetorotational instability
(MRI) in protoplanetary discs can account
for the wide diversity of observed
accretion outbursts. We show that
short-lived, relatively low accretion rate events
probably result from triggering in the inner disc
and can occur at low surface densities, comparable
to or smaller than the minimum mass solar nebula,
and thus are very unlikely to result from MRI triggering
by gravitational instability. We develop time-dependent accretion disc models using an $\alpha$-viscosity approach 
and calculate light curves to compare with observations.
Our modeling indicates that
the lag time between infrared and optical bursts seen in Gaia 17bpi can be explained with an outside-in
propagation with an
$\alpha \sim 0.1$ in the MRI-active region,
consistent with other estimates.
While outbursts in inner discs can show time delays of a few years
between infrared and optical light curves, our models 
indicate that large, FU Ori-like bursts can exhibit infrared precursors decades before optical
bursts. Detecting such precursors could
enable analysis of the
central star before it is overwhelmed by the rapid accreting material, as well as
constraining 
outburst physics. Our results emphasize the importance
of near-infrared monitoring of young stellar objects in
addition to optical surveys.
In addition, our findings emphasize
the need
for more sophisticated, three-dimensional, non-ideal magnetohydrodynamic simulations to fully exploit
observational results.
\end{abstract}


\section{Introduction}

Potential or actual planet-forming discs accrete at variable rates onto their young central stars 
\citep[see references in][]{hartmann16}. 
In some cases accretion rates can change by orders of magnitude over timescales of months to years.
Such 
accretion events or outbursts
can provide
insights into mechanisms of mass transport that are otherwise difficult to constrain.  Outbursts also can result in significant thermal processing of protoplanetary material
\citep{bell00,ciesla07,ciesla10}.
The amount of material transported in
accretion bursts also indicates
lower limits to mass reservoirs of inner discs. 

The largest accretion outbursts 
are those of the FU Orionis variables
\citep{herbig77,hartmann96},
that can involve accretion at rates $\sim 10^{-5} - 10^{-4} \msunyr$ lasting tens of years up to 100 years or more.  While such outbursts are relatively rare, 
the advent of systematic wide-field
photometric surveys have turned up increasing numbers of FUors \citep[see][for reviews]{audard14,fischer22}.
Several different mechanisms have been proposed to explain FUor outbursts; accretion from a disc
that causes the central star to expand
\citep{larson83},
fragments resulting from gravitational instability (GI)
\citep{vorobyov06,vorobyov10},
{ disruption of young massive planets
\citep{nayakshin12,nayakshin23}, encounters with companion stars \citep{forgan10}, planet-disc interaction \citep{lodato04}, and
GI-driven turbulence that triggers the magnetorotational instability \citep[MRI;][]{armitage01,zhu07,zhu09,zhu10a,zhu10b,bae13,bae14,kadam20}
\citep[see][for a discussion]{vorobyov21}.

Mass and angular momentum transport
in pre-main sequence discs is generally thought to be slow.  The low ionization states thought to be characteristic of these discs is believed to 
preclude efficient transport by the MRI.
Observations also indicate that levels of turbulence, whether hydrodynamic or magnetic, are very low 
(at least in outer discs; \citep{flaherty15,flaherty20},
 consistent with slow transport rates.  Models of magnetic wind-driven accretion
 \citep{baistone13a,baistone13b,bai6} do predict rapid global
accretion flows, but these are confined to thin upper layers of the disc, only draining the bulk of the
mass over long timescales.
However, if for some reason the disc
temperatures can rise above $\sim 1000-1200~$K, the rapid increase in thermal ionization can result in activating the MRI
\citep{desch15}, with a corresponding large increase in angular momentum transport that results in an accretion 
outburst.  In support of 
this picture,  
the FU Ori objects show optical and near-infrared spectra of gaseous
photospheres
with local effective temperatures
$\gtrsim 2000$~K, easily large enough to produce sufficient thermal ionization
for the MRI to operate.


Photometric surveys are increasingly 
finding outbursting objects with smaller accretion rates than in the canonical FU Ori objects \citep[e.g.,][]{contreras14,lucas17,park21,hillenbrand18,contreras23}.
Such events may be difficult to explain with triggering by 
gravitational turbulence,  as GI requires high surface densities that tend to produce high accretion rates.  Nevertheless,
the spectral signatures of many of these smaller bursts (e.g.,
optical effective temperatures) are similar to
those of FUors, indicating that activation of the MRI is
likely. 

Recent observations of two outbursting
events, Gaia 17bpi \citep{hillenbrand18,rodriguez22},
and Gaia 18dvy \citep{szegedi20},
are of particular interest in that
the outbursts clearly start earlier at
near-infrared wavelengths than in the
optical regime. This behavior shows that
the outbursts are propagating outside-in,
with the increase in temperature appearing
first at large radii where the discs
are cooler. In principle, the time lapse
between the optical and infrared light
curves yield constraints on the propagation
speed of the outburst and other physical
parameters.

These recent observational developments motivate this effort
to compute outburst light curves
using time-dependent simulations
in a preliminary exploration of
what might be learned from present
and future observations.

\section{Overview of conditions for MRI activation}

\label{sec:overview}

To set the stage for understanding the behavior
of the numerical simulations,
and the resultant outburst light curves,
we outline some basic
considerations.
Activation of the MRI by thermal
ionization in the
outer disc requires
temperatures well in excess
of those attainable either
by
irradiation heating from the
central star or internal viscous heating at typical
T Tauri accretion rates.
For a steady disc the
effective temperature 
resulting from viscous heating well 
beyond the inner radius is
\begin{equation}
T_e \sim 60 M_{0.3}^{1/4} \mdot_{-8}^{1/4}
R_{au}^{-3/4}\,,
\end{equation}
where $M_{0.3}, \mdot_{-8}$,
and $
R_{au}$ are the stellar mass in units of $0.3 \msun$, the accretion rate in units of
$10^{-8} \msunyr$, and the radius in au.
Even during a massive
outburst with $\mdot = 10^{-4} \msunyr$, the effective temperature at 1 au would only 
be $\sim 600$~K. Thus activating and maintaining
the MRI requires the central
temperature to be larger than
that of the surface, especially from the
quiescent state at typical
T Tauri accretion rates
of $\sim 10^{-8} \msunyr$.
Models that can achieve this
require radiative trapping,
which in turn requires
the disc to be significantly
optically-thick
\citep{armitage01,zhu09,zhu10a,zhu10b}.

To fix ideas, consider a simple
model along the lines
of \cite{zhu10a,zhu10b}. We suppose that some
mechanism produces a local jump in the temperature, high enough to make
a transition from the low-viscosity initial state to an MRI-active
state. 
 The resulting viscous heating is \citep[e.g.,][]{bell94}
\begin{equation}
    F_V = {9 \over 4} \alpha \Omega c_s^2 \Sigma\,.
    \label{eq:fviscous}
\end{equation}
Here we employ the ``alpha'' viscosity formalism, where $\Omega$ is
the Keplerian angular frequency,
$c_s$ is the (midplane)
sound speed, and $\Sigma$ is the surface density.
We also assume that this increase in energy generation is instantaneously
balanced by radiative losses of the optically-thick disc with effective
temperature $T_e$,
\begin{equation}
F_V = 2 \sigma T_e^4\,,
\label{eq:teff}
\end{equation}
with a factor of two to account for both sides of the disc.  
Radiative trapping raises the central temperature
$T_c$ at large optical depth relative to the effective
temperature $T_e$ such that
\citep{hubeny90,zhu10a}
\begin{equation}
    T_e^4 = {8 \over 3} {T_c^4 \over \tau_R}\,,
    \label{eq:te_tc}
\end{equation}
where $\tau_R = \kappa_R \Sigma$ is the Rosseland mean optical depth.
(Here we ignore heating by irradiation from the central star for simplicity.)

While the Rosseland mean opacity $\kappa_R$ generally depends upon temperature and density,
here we only establish a critical surface density at a particular
activation temperature. Following the results of
\citep{desch15} that indicate activation temperatures lower than those that would result in sublimation of grains, we assume that the opacity is
dominated by dust. This allows us to simplify the solution by adopting
a constant value for $k_R$, since the temperature and density dependences
are weak \citep{zhu09}.
As shown later in discussing FU Ori-type large bursts, our results are broadly consistent with the following numerical simulations and
with our previous calculations that explicitly use the temperature and density dependence of the opacity \citep{zhu10a,zhu10b,bae13}.

Substituting for $T_e$ on the right-hand side of equation \ref{eq:teff}
using equation \ref{eq:te_tc}, setting $T_c = T_i$, the adopted activation temperature, and
solving for the surface density results in
\begin{equation}
\Sigma_i^2 = { 64 \sigma \over 27} {\mu \over \alpha \Omega \mathcal{R}} {T_i^3 \over \kappa_R}\,,
\label{eq:sigmacrit}
\end{equation}
where $\mu$ is the mean molecular weight
and $\mathcal{R}$ is the gas constant. This is the limiting surface density needed to raise the central temperature to $T_i$, given $\alpha$, $\Omega$, and $k_R$.
For fixed values of $\alpha$ and $\kappa$, $\Sigma_i \propto R^{3/4}$.  Thus, it is easier to trigger the MRI at smaller radii, as temperatures are naturally higher. 

Thus, equation \ref{eq:sigmacrit} indicates
that there will be a minimum
value of the surface density
at which the MRI can be
sustained, with
$\Sigma_c \propto T_i^{3/2}
\alpha^{-1/2}
\kappa^{-1/2} R^{3/2}$ 
for a given central star mass
$M$. The numerical simulations
presented in the following
sections exhibit this basic
scaling, although with
significantly larger values. This is because this 
steady-state model does not account for
time-dependent changes in $\Sigma$,
as discussed in the following section.

\section{Time-dependent outburst models}

\subsection{Methods}

\label{sec:methods}

A full treatment of triggering the MRI in
a protoplanetary disc would require three-dimensional, non-ideal magnetohydrodynamic (MHD) calculations in a resistive disc with
radiative transfer. To our knowledge such calculations have not yet been conducted and would
in any case be be subject to considerable
uncertainties (for example, the properties and
spatial distribution of dust). Given these
complexities, a simpler approach is warranted
to outline basic issues.

We computed
time-dependent 1-dimensional disc models using
the equations and methods described in
\cite{zhu10a}, with a few small modifications.
The disc has an MRI-active layer with surface density
$\Sigma = 10\, \gmcmtwo$, or is otherwise
non-viscous, with a central "dead zone" unless
the temperature is raised above some fiducial
value. In this case we turn on the viscosity $\alpha$ parameter to
its full value smoothly as the
temperature rises from 1200~K to 1300~K,
on an e-folding timescale of $2 \Omega^{-1}$,
where $\Omega$ is the Keplerian angular frequency
(see equation 17 in \cite{zhu10a}), but without
the second term for radial diffusion of the magnetic
field; we find that that
term makes essentially no
difference. As discussed in the Appendix, modest changes in the activation temperature do not significantly change results.
The active layer plays essentially no role
in these calculations other than setting
a base low accretion rate $\mdot \lesssim 10^{-9} \msunyr$,
which does not not have a significant impact on the outburst light curves.
Most models were calculated with 256 grid steps over ranges from
$\sim 0.1$ to 10 au, with a few test models run with 512 radial steps.

We assume that when the MRI is activated it
results in a maximum viscosity parameter $\alpha = 10^{-1}$. This choice is motivated by results for Gaia 17bpi that are strongly inconsistent with observations assuming
$\alpha = 0.3$ and $= 0.03$ (see Appendix).

We adopt the opacities given by \cite{zhu09},
except we reduce the opacities below 1400~K
by a factor of ten to simulate depletion
of small grains. Unlike previous papers
\citep{zhu10a,zhu10b,bae13,bae14}
there is no infall to the disc. In addition, we weaken
the transport due to gravitational instability 
by setting the viscosity due to GI as
$\alpha_Q = 0.05 \exp(-Q^4)$, where 
$Q = c_s \Omega /(\pi G \Sigma)$ is the Toomre
parameter. This was done to focus on the conditions needed
for MRI-triggering in more generality than using GI heating.

We impose a high central
temperature perturbation at a given radius, 
as we are agnostic about the source of this energy
release. 
A top-hat temperature increase to $T= 1500$K in a specified
radial range is assumed. Whether or not an outburst occurs depends
mostly on the magnitude of the surface density
$\Sigma$ at the position of the temperature increase
and to some extent on the radial width of the
perturbation.  These details depend upon our
assumption of a constant $\alpha$ among other
things; as we are most interested in the
light curves assuming the MRI is triggered,
these details are unlikely to change the
main results.

Our initial surface density distributions are for
the most part assumed to be flat inside the radius
of the temperature perturbation, with a decline at
longer radii. The choice of an inner flat
$\Sigma(R)$ is motivated by several reasons,
beyond simplicity:
1) Our results in \cite{hartmann18} for viscous disc accretion at low values of $\alpha$ lead to fairly
flat density distributions (as illustrated in
\S \ref{sec:sigmacrit}). (2) Our simulations for
FU Ori objects show that the inner disc becomes depleted
after the first outburst \citep{zhu10b,bae13}.
The detailed $\Sigma(R)$ is not completely flat, but experiments indicate that outburst properties are not sensitive to modest (factor of 2-3) 
variations.
(3) Simulations with steeply increasing $\Sigma(R)$ 
are prone to generate inside-out bursts, in contrast
to most observations (see following discussion).
(4) The choice of a flat density distribution does
not affect the outburst, but minimizes the peak
accretion rate, as discs with more mass at smaller
radii accrete more.  On the other hand,
a decreasing surface density with radius in the
outer disc is adopted to avoid GI-triggering, which
we have already explored for FU Ori outbursts
\citep{zhu10b,bae13}.

\begin{figure*}
\includegraphics[width=1.0\textwidth]{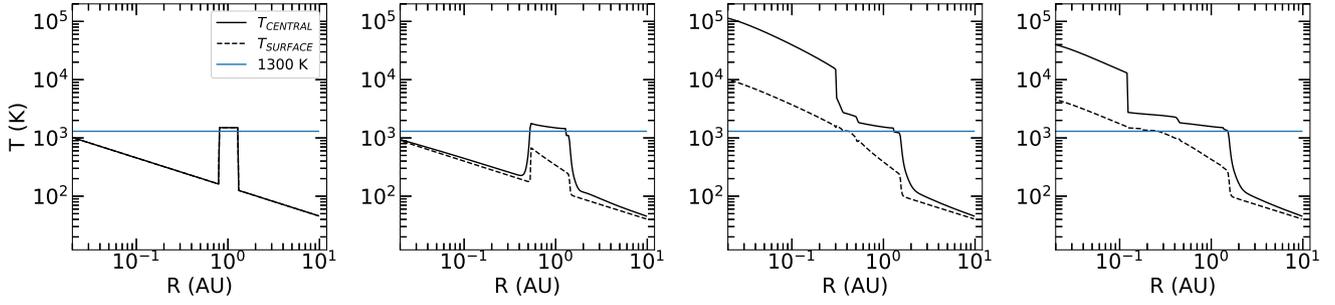}
\caption{The central temperature (solid line) and the surface temperature (dashed line) at $t = 0$, 125, 175, and 
183 years (compare with Figures \ref{fig:dec82_sigmas} and \ref{fig:dec82_mdot}).
The horizontal line denotes
$T = 1300$~K. The large increase in temperature
seen in the last two panels results from
thermal instability (see text).}
\label{fig:dec82_temps}
\end{figure*}

\begin{figure*}
\includegraphics[width=1.0\textwidth]{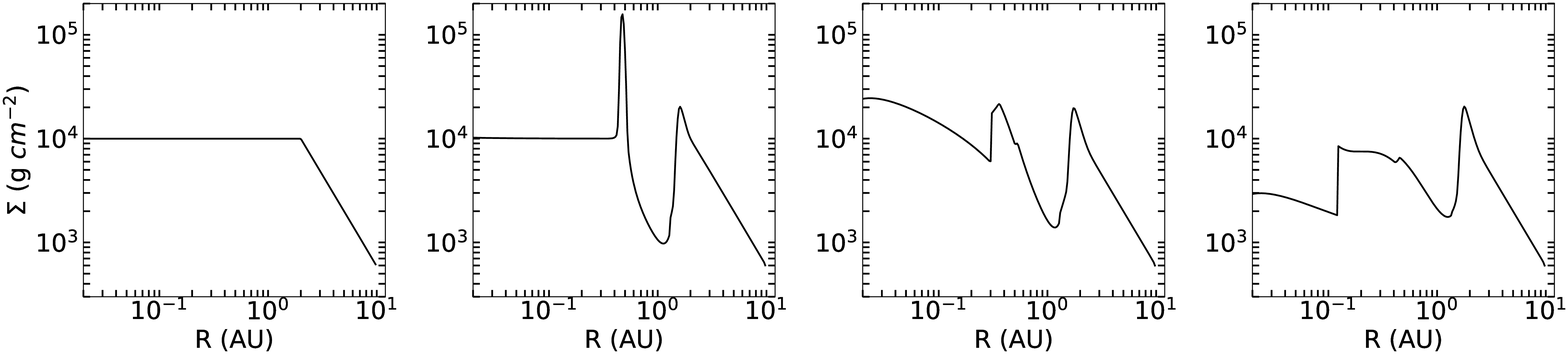}
\caption{Surface densities at the same times as in Figure \ref{fig:dec82_temps}. 
corresponding to times $t = 0$, 125, 175, and 
183 years (compare with Figure \ref{fig:dec82_mdot}). The activation of the
MRI, and the resulting jump in viscosity, causes
matter to move inward and outward (to conserve
angular momentum), resulting in a depletion of mass
at the site of the original perturbation.}
\label{fig:dec82_sigmas}
\end{figure*}

To calculate the optical spectrum it is necessary to
capture regions where the disc is hottest. This means
we must adopt small innermost radii $= 0.01 - 0.015$~au,
much smaller than
typically used in our previous calculations
\citep{zhu10b,bae13,bae14} and in other simulations
\citep{vorobyov06,vorobyov10,vorobyov20,kadam20}.\footnote{We note that \cite{gehrig22} did consider small inner radii but concentrated on the interaction with the stellar magnetosphere and consequent evolution of stellar rotation, not the
resulting spectra.}
If there is a significant surface density at
such radii, the heating of the disc by irradiation from the central star will make the innermost radii MRI-active
\citep[e.g.][]{gehrig22}.
This can result in oscillating behavior at the
transition between active and inactive regions
\citep[see, for instance, the rapid oscillations
in accretion rate see at early times in
Figures 2 and 3 in][]{bae13}.

More generally, the activation of the MRI at the inner
disc edge produces an inside-out burst.
 This is not a difficulty for models for which the main temperature perturbation is taken to be at large radii, because the initial inside-out
event can die out first. However,
models comparable to the observations of
Gaia 17bpi need to be perturbed at small radii to produce
the necessary short timescales,
and it is easy for the inside-out burst, that would exhibit little to no lag between infrared and optical light curves, to wipe
out the outside-in outburst. The structure of the inner region is further complicated
by the likely presence of the stellar magnetosphere, which typically truncates the
disc at 3-5 stellar radii \citep{bouvier20,garcia20}. With stellar radii
$\sim 1-2 \rsun$, this implies disc truncation
at about $0.015 - 0.025$~au ($\sim 3-5 \rsun$).
As our interest is in outbursts, we avoid constructing a self-consistent steady inner disc
structure, by adopting a low irradiation temperature
$T_{ext} = 1200 (r/ 0.15 au)^{-1/2}$
and adopt a low surface density inside
of $\sim 0.02$~au, depending on the model.
This means that the pre-outburst accretion rates
are not reliable, but as we are interested in
the behavior of outbursts
the very innermost disc pre-burst state makes little difference to the burst light curves.

\subsection{Basic behavior}

\label{sec:basic}

In analyzing the model results
we did not conduct a broad search of
parameter space, because the accuracy with which
our simple $\alpha$ viscosity models can represent
non-ideal MHD results is uncertain.
Instead, we focus on broad
trends that hopefully can ultimately
be tested by observations.

The basic behavior of our
outburst models are
broadly similar to those presented
in earlier papers \citep{zhu10a,zhu10b,bae13,bae14}.
Figures \ref{fig:dec82_temps} and \ref{fig:dec82_sigmas}
show snapshots of the disc temperature and surface density distributions at selected times assuming a temperature perturbation at 1 au at 
$t = 0$. During the
initial propagation of the
outburst (second panels,
$t = 125$~yr), the large increase in $\alpha$ in the
central regions from very low initial values causes mass to 
both move inward in a dense
front, while moving outward at larger radii to conserve
angular momentum. This results in a surface density minimum near the site of the
original perturbation. 

The accretion wave tends to
move inward at a temperature
$\sim 1500$~K; this is a thermostatic effect of the
assumed evaporation of dust
\citep[see, e.g][]{dalessio01}, assuming the
\cite{zhu09} opacity.
However, once the wave
moves inward of about
0.5 au, the temperature makes
a dramatic jump to very high
values. This is due to the thermal
instability that occurs once
the dust is gone, 
because of the steep temperature
dependence of the gas opacity
\citep{bell94}. Note that though the central
temperature exhibits jumps, the surface
(effective) temperature closely follows the
$T \propto R^{-3/4}$ distribution characteristic
of the optically-thick steady disc.

The accretion front does
not reach the inner edge of
the disc until $t \sim 175$~yr
(Figure \ref{fig:dec82_mdot}).
The high surface density 
and high viscosity of the
hot inner region results in
a large, fast jump in the
accretion rate.
 There are also
brief ``drop-outs'' in $\mdot$  as the
outburst proceeds. The reason
for the accretion dips is
understandable from an
inspection of the final panels in
Figures \ref{fig:dec82_temps}
and \ref{fig:dec82_sigmas}
at $t= 183$~yr. The inner disc surface density at this time has decreased by an order of magnitude from the preceding
snapshot at $t=175$~yr.
Because the temperature of
the inner disc is much higher
than that of the outer disc,
and the viscosity is assumed to scale as $\nu \propto \alpha T$, 
the mass transport is much
faster in the inner hot region than in the outer, cooler zone.
This results in
the depletion of mass in
the hot inner region of mass 
until transport from the cooler region can replenish the material.
In essence, this is a version
of the ``S-curve'' thermal
instability mechanism for repeated outbursts \citep[see][and references therein]{bell94}.

\begin{figure}
\includegraphics[width=0.45\textwidth]{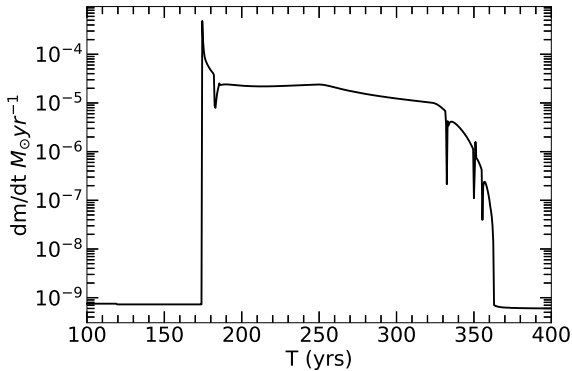}
\caption{Accretion rate at the inner disc edge as a function for the model shown in
Figures \ref{fig:dec82_temps} and \ref{fig:dec82_sigmas}.}
\label{fig:dec82_mdot}
\end{figure}

Rapid, temporary dips in
brightness have been observed
in the FU Ori objects V1057 Cyg and
V1515 Cyg \citep{kolotilov97,kenyon91,kopatskaya13,szabo22}, with perhaps a longer,
somewhat different drop seen in V2493 Cyg \citep{semkov21}. Whether these
observed dips are the result
of this mechanism, instead of
dust extinction events, is
not clear. If these are do to an accretion
drop-out, one would expect the optical spectrum to have a later
spectral type (cooler maximum
terperature) during
minimum, whereas a dust extinction event would not show a change. Detailed spectral and photometric monitoring
would be needed to test this.
In any case, it should be emphasized that our 1D models assume axisymmetry, which is
unlikely in MHD turbulence, and
thus may result in a much bigger
effect than in more complex,
non-axisymmetric MRI behavior, including convection \citep{zhu10a}.

\subsection{Critical surface density}

\label{sec:sigmacrit}

We constructed a set of outburst models triggered at
several different radii.
At each position we tried different initial surface densities to find the minimum
value of $\Sigma$ for which
the perturbation could be sustained and led to an outburst.
The central star mass in all cases was $0.3 \msun$ and we adopted
the irradiation temperature and
flat surface density distribution interior
to the position of the perturbation 
as discussed in
\S \ref{sec:methods}, 
except with an inner ``magnetospheric' depletion as needed. 
The results are
shown by the circles in Figure \ref{fig:sigmas}.

We can use the developments of
\S \ref{sec:overview} to understand
the dependence of the critical
surface density on radius, but with
an important difference. Using 
equation \ref{eq:sigmacrit} and adopting the dust opacity from \cite{zhu09} reduced by
an order of magnitude to account for some depletion of small dust, results in
\begin{equation}
    \Sigma_i \sim 1 \times 10^3
    (\alpha_{-1} M_{0.3}\,\kappa_i)^{-1/2}
    R_{au}^{3/4} T_i^{3/2}\,
    {\gmcmtwo}\,,
    \label{eq:sigmacrit_num}
\end{equation}
where $\kappa_i$ in units of $0.87 {\, \rm cm^2}  g^{-1}$, $T_i$ is in units of $1300$~K,
the stellar mass is $M_{0.3}$ is in units of
$0.3 \msun$, $R_{au}$ is self-evident, and $\alpha_1$ is in units
of 0.1, so that the scaling is appropriate
for our simulations.

In Figure \ref{fig:sigmas} we show 
that while the prediction of equation \ref{eq:sigmacrit_num} exhibits the
correct dependence on radius, it
underpredicts the limiting surface
densities in the simulations by a factor
of $\sim 10$. After considerable 
exploration it appears that the
reason for the discrepancy is the
change in $\Sigma$ due to huge increase
in viscosity by the applied temperature
perturbation. As shown in the second panel of
Figure \ref{fig:dec82_sigmas},
the activation of the $\alpha$
viscosity generates a large drop in
$\Sigma$ at the central position of
the initial perturbation as mass
diffuses away both inward and outward.
The magnitude of this decrease in 
surface density is typically a factor
of 10, such that this minimum must be close to,
or larger than,
the value of $\Sigma_c$ from
equation \ref{eq:sigmacrit_num} for
the outburst to proceed rather than die away. 
As a specific example, the second panel of Figure \ref{fig:dec82_sigmas} 
shows a minimum value of $\Sigma \sim 1 \times 10^3 {\rm g \, cm^{-2}}$,
in agreement with equation \ref{eq:sigmacrit_num} at 1 au.
Equation \ref{eq:sigmacrit_num} indicates
how $\Sigma_c$ would change due to differing
adopted parameters other than the radius.

\begin{figure}
    \centering
    \includegraphics[width=0.45\textwidth]{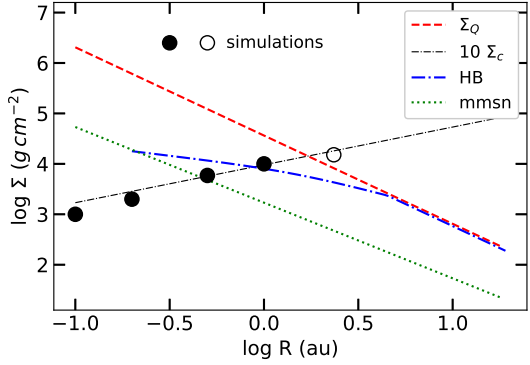}

    \caption[Limiting surface densities]{Limiting surface densities for outbursts (filled circles) for the standard set of numerical simulations.  The open circle denotes a model where the
    model gravitational viscosity is important.
    For comparison we show 10 times the "critical" surface density derived in equation
    \ref{eq:sigmacrit}, along
    with 
    the limiting value for gravitational instability 
    $\Sigma_Q$ (dashed line) as an upper bound, the minimum 
    mass solar nebula surface density (dotted line), and a 
    model for low-state 
    accretion at $2 \times 10^{-8} \msunyr$ 
    and $\alpha = 3 \times 10^{-4}$ 
    from \cite{hartmann18}. }
    \label{fig:sigmas}
\end{figure}



We compare the results 
in Figure \ref{fig:sigmas}
with three
different surface density distributions:
the limiting value of the surface density for gravitational instability setting 
$Q = c_s \Omega/ (\pi G \Sigma_Q) = 2$, assuming
$T_{irr} = 200 (R/au)^{-1/2}$~K (red dashed line); the canonical minimum-mass solar nebula (green dotted line);
and the
surface density of a model by
\cite{hartmann18} consistent with
observed T Tauri accretion rates
$\sim 10^{-8} \msunyr$ for a low state
$\alpha = 3 \times 10^{-4}$.
The comparisons indicate that much
lower surface densities can suffice
to sustain an MRI-driven outburst at small radii.
Conversely, large surface densities
are required to trigger the MRI at
large radii where the disc is more
likely to be gravitationally unstable.
Sustaining an MRI outburst
at radii greater than a few au
would require surface densities far
above the limit for gravitational
stability, so this predicts an outer limit for
the high-temperature region.

\begin{figure}
\includegraphics[width=0.45\textwidth]{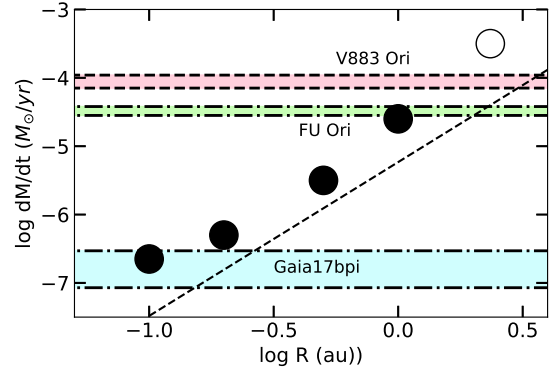}
\caption[dependence of accretion rate]{Dependence of the central
mass accretion rate, estimated
at the ``plateau'' value immediately
after the main spike (see Figure \ref{fig:dec82_mdot}), as a function
of radius for the model sequence
displayed in Figure \ref{fig:sigmas}.
The dashed line is given by 
equation \ref{eq:mdoti}.
We use large dots to indicate that the precise values depend on exactly the time at which the (varying) accretion rate is measured.
The ranges of estimated accretion rates for
Gaia 17 bpi are taken from \cite{rodriguez22} (RH22), but
multiplied by a factor of two to account
for the difference between their assumption of $0.6 \msun$ for the central mass in contrast to the $0.3 \msun$ of
the present calculation (the disc luminosity and temperature depend upon
the joint product of $M_* \mdot$.
The ranges for FU Ori and V883 Ori are the
best values from RH22 and 
\cite{liu22}; these probably underestimate
the true uncertainties but are used to avoid
overlap (see discussion in \S 4).}
\label{fig:mdotr}
\end{figure}

\subsection{Accretion rates and decay timescales}

\label{sec:accdecay}

The simulations unsurprisingly predict that larger, longer
outbursts result from triggering at larger radii.
Figure \ref{fig:mdotr} shows the model sequence for the
triggering radii, assuming 
minimum surface densities, in comparison with observational
estimates of mass accretion rates for several objects.
Using the new Gaia distance of 416 pc and adopting an inclination of $i = 35^{\circ}$, \cite{zhu20} estimated an accretion rate of $\sim 4 \times 10^{-5} \msunyr$ for
FU Ori. Note that this is a current estimate, as the object has faded
at optical wavelengths by more than a magnitude \citep[e.g.,][]{kenyon00}. We take a range of $2.8 -3.8 \times 10^{-5} \msunyr$
for FU Ori and
$7 - 11 \times 10^{-5} \msunyr$ for V1883 Ori from
\cite{liu22}. For Gaia 17bpi,
\cite{rodriguez22} (
RH22) estimated $\mdot \sim 0.9 - 6 \times 10^{-7} \msunyr$. Then Figure \ref{fig:mdotr} suggests triggering
at $\sim 2 \sim 1$, and $\sim 0.1$~au for V1883 Ori, FU Ori,
and Gaia 17bpi, respectively.

Using equation \ref{eq:sigmacrit_num},
one can make a crude estimate of
the expected mass accretion rate
using the formula for a steady disc,
\begin{eqnarray}
    \mdot_i & = & 3 \pi \alpha c_s^2 \Sigma \Omega^{-1} \\
    &=& 
    2.2 \times 10^{-5}
    (\alpha_{-1} \kappa_i)^{1/2} M_{0.3}^{-3/4}
    R_t^{9/4} T_i^{5/2}
    \msunyr \,,
\label{eq:mdoti}
\end{eqnarray}
where $R_t$ is the radius in au at which the perturbation was placed.
Figure \ref{fig:mdotr} shows that equation \ref{eq:mdoti} again reproduces the radial scaling, but lower than predicted by
equation \ref{eq:mdoti}; this is as expected from the results for the critical
$\Sigma$ (\S \ref{sec:sigmacrit}.
These are minimum accretion rates due to the assumption of
a flat inner surface density distribution; $\mdot$ can
be a few times higher  for models
with increasing surface densities with decreasing radius,
as discussed later in \S \ref{sec:fuori}.

\begin{figure}
\includegraphics[width=0.45\textwidth]{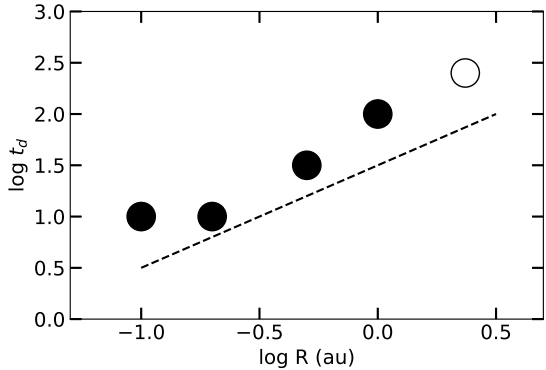}
\caption{Approximate decay timescales of the central mass accretion rates for
the models, defined as the time taken
to decay by a factor of two from the
plateau immediately after the initial
spike in $\mdot$, as a function of the
radius at which the initial perturbation
was applied. Using the entire
length of the outburst instead leads
to similar results, with timescales
larger by about 30-40\%. The dashed line denotes
$t_d \propto R$ for comparison.}
\label{fig:tdecay_r}
\end{figure}

Figure \ref{fig:tdecay_r} shows decay timescales
for the models. Relating this simply to a viscous time is
complicated because of the
nature of the central temperature distribution, with a hot inner region and a colder outer disc, although one expects this to depend approximately inversely with
$\alpha$. As indicated by the dashed line, the dependence of
the decay time on radius is
found to be roughly proportional to $R$.

In the one-dimensional outburst models by \cite{zhu10a,bae13}, mass infall to the outer disc is transported inward by gravitational instabilites, eventually resulting in enough heating to trigger the MRI at a
radius $\sim 3$~au; The results in Figure \ref{fig:mdotr} are roughly consistent
with this triggering radius. The observed rate
of decay of FU Ori, $\sim 0.015 \rm{\, mag \, yr^{-1}}$
in the B and V bands \citep{kenyon00}, is reasonably
consistent with the predicted $\sim 100$~yr decay
timescale (Figure \ref{fig:tdecay_r}).

\section{Disc evolution and light curves}

\label{sec:lightcurves}

A major motivation in constructing simple time-dependent accretion models is to make comparisons with observed
light curves.
Following RH22, the spectral energy distribution for each model at each snapshot in time (taken at 1/2 year
intervals)
was calculated from the sum of the spectra
of each annulus taken from
a model atmosphere with the same effective temperature.
We used the
BT-NextGen\footnote{\url{http://svo2.cab.inta-csic.es/theory/newov2/index.php?models=bt-nextgen-agss2009}} (AGSS2009) calculations \citep{allard11,allard12} for
temperatures between 15,000 K and 2700 K,
setting $\log g = 1$ as appropriate
for discs \citep[see, e.g.][]{liu22}.
For lower temperatures between 1400 and 2700 K we use the BT-Settl\footnote{\url{http://svo2.cab.inta-csic.es/theory/newov2/index.php?models=bt-settl-agss}} (AGSS2009) models. For these temperatures we take the lowest surface gravity available
in the model set, $\log g = 3.5$ \citep{liu22}. At temperatures below
1400 K we assume that dust forms a featureless continuum and that the
resulting spectrum is that of a blackbody.
In all cases we assume an inclination
of $60^{\circ}$ to the line of sight.

We found it unnecessary to keep the full (very high) spectral resolution of the atmosphere models. Following RH22,
We include the flux of a central star of a given
radius and effective temperature that radiates as the corresponding BT-NextGen model.

\begin{figure*}
    \centering
    \includegraphics[width=0.9\textwidth]{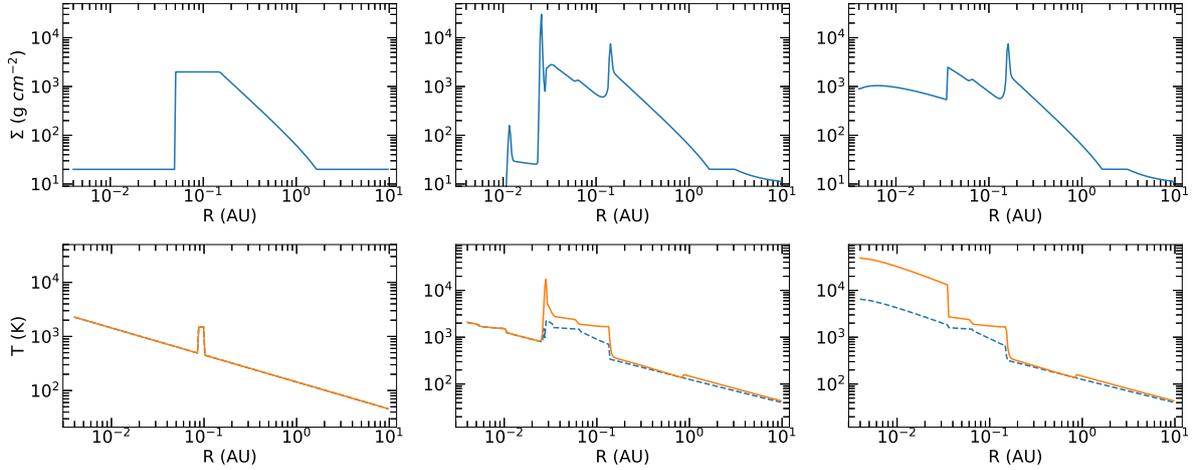}
    \caption{Evolution of an outburst model to compare with observations of Gaia17bpi.  The upper panels show the the surface density structure and temperature at three times; the initial condition, the development of the outburst, and the structure at peak accretion rate. The dashed and solid curves show the surface (effective) temperature and the central temperature, respectively. From left to right, the plots outline structure occurring at 0, 4, and 5 years from the start of the model.} \label{fig:mar1structure}
\end{figure*}

For comparison with observations that are
reddened by interstellar extinction, we
used the PyAstronomy module Unred\footnote{\url{https://pyastronomy.readthedocs.io/en/latest/pyaslDoc/aslDoc/unredDoc.html}} with the extinction correction for $R_V = 3.1$ as defined in \cite{Fitzpatrick99}. Supplying in a negative value for the color excess E(B-V), we were then able to add in wavelength-dependent extinction.

To facilitate comparison with observations,
we computed magnitudes for the GAIA G, WISE W1, WISE W2, CTIO V, and the 2MASS J, H, and K filters
\citep{cutri03}.
We obtained transmission-wavelength data from the SVO Filter Profile Service\footnote{\url{http://svo2.cab.inta-csic.es/theory/fps/}} as defined in \citep{Rodrigo12, Rodrigo20}. 
The GAIA G magnitudes were computed with the method presented by \cite{Evans18}.

In the following we compute light curves and 
spectral energy distributions (SEDs) of models exhibiting small
and large outbursts. We compare the
results with observations of
Gaia 17bpi and FU Ori that may be taken as representative of these two cases. Our goal is not to model the data precisely, which is not possible
in any case given the limitations of the simulations, but to see the extent to which simple models can explain the overall outburst
behaviors.

\subsection{Weak outburst: Gaia 17bpi}

\label{sec:gaia}

For purposes of comparison to the
observed light curves of Gaia 17bpi
and to previous modeling, we fix the distance
as 1.27 kpc \citep{hillenbrand18} and adopt the visual extinction
of $A_V = 3.5$ as estimated by RH22.
The high-state accretion rate of $\sim 1 - 3 \times 10^{-7} \msunyr$ estimated
by RH22 assumed a central mass of
$\sim 0.6 \msun$. For these models,
where we have instead assumed $M_* = 0.3 \msun$, the corresponding accretion
rate would be $\sim 2-6 \times 10^{-7} \times \msunyr$, because
the product $\mdot M_*$ enters in to both the luminosity and maximum disc temperature 
of a steady optically-thick disc.
For the final model we assume
an inner radius of 0.003 au to correspond more closely to the (very small) stellar radius of 
$\sim 0.4 \rsun$ of RH22. 
(We had numerical problems going to smaller
radii.) A viscosity
parameter of $\alpha = 0.1$ was assumed.
RH22 assumed a central star
mass of $0.6 \msun$, but we find no
significant difference in the outburst
changing mass by a factor of two.

Figure \ref{fig:mdotr} indicates that the temperature
perturbation should be located interior
to about $0.2$~au, and that a minimum
surface density should be $\sim 2000 \, \gmcmtwo$. Consistent with
these expectations, the initial condition of our
final model shown in the left panel of
Figure \ref{fig:mar1structure}
exhibits a temperature perturbation
centered at 0.09 au at a region of
constant surface density
$\Sigma = 2000 \, \gmcmtwo$. The surface
density is strongly depleted
inside of 0.05 au to avoid
having the innermost regions go into outburst
first, resulting in an outside-in burst
(\S \ref{sec:methods}). We interpret our evacuated
inner region as corresponding to magnetospheric truncation.
The falloff of $\Sigma$ beyond 1.5 au
was introduced in a (failed) attempt to make
the WISE fluxes smaller, but it has little effect on the outburst.
(Further discussion of how we arrived
at this model, along with illustrations
of the effect of changing various 
parameters, is presented in the Appendix.)

The disc evolution shown
in Figure \ref{fig:mar1structure}
is similar to that shown
in Figures \ref{fig:dec82_temps} and
\ref{fig:dec82_sigmas}. The onset of a high viscosity results in the propagation
of mass inward in a sharp spike (see
middle panel), along
with an outward pileup to conserve
angular momentum, and a significant
minimum in the surface density in between.
During this phase the outburst would be detected only at infrared wavelengths.
As the accretion front reaches the star,
the central temperature jumps up and
reaches very high values due
to thermal instability (see right panel). 

As shown in
Figure \ref{fig:mar1_mdot}, the accretion
rate plateaus at a value $\sim 3 \times 10^{-7} \msunyr$, as expected.
The peak temperature in the inner disc
is $\sim 6000$~K, somewhat cooler than
the $\sim 7600$~K estimated by RH22.
We note that our models employ a flow-through inner boundary condition, and so they do not show the turnover of the effective temperature that results in steady disc models with a zero-torque boundary condition.

\begin{figure}
    \centering
    \includegraphics[width=0.45\textwidth]{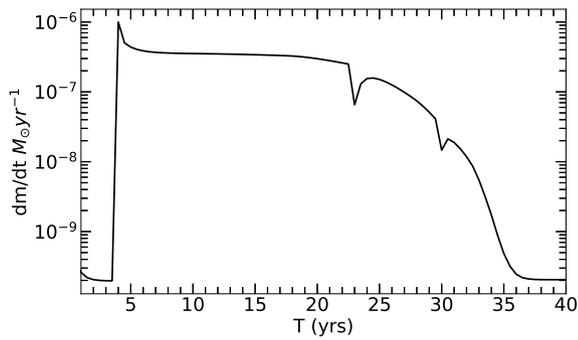}
    \caption{Central accretion rate vs. time for the model shown in Figure \ref{fig:mar1structure}.}
    \label{fig:mar1_mdot}
\end{figure}

\begin{figure}
   \centering
   \includegraphics[width=0.45\textwidth]{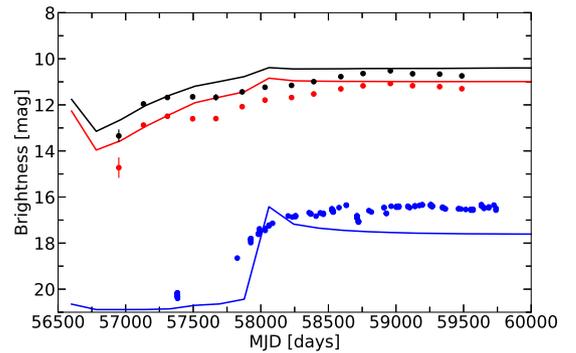}
   \caption{Model light curves resulting from
   the initial conditions shown in Figure \ref{fig:mar1structure},
   compared with photometry of Gaia 17bpi. The black and 
   red dots and curves are the WISE W1 and W2 magnitudes and model results, respectively, while
while the blue dots and curve are the Gaia G band photometry and corresponding model magnitudes. The time $t = 0$
in Figure \ref{fig:mar1_mdot} corresponds to MJD $=$ 56600. 
The initial drop in the
infrared light curves is due to transient accretion at the inner edge of the disk that dies away rapidly (see text).
The rise time of the optical light curve is unresolved at
the plotted 0.5 year sampling. A test run with higher resolution
sampling indicates that the true rise is about 1 month (the individual time steps of the calculation are much shorter).
}
   \label{fig:mar1.lightcurve}
\end{figure}

Figure \ref{fig:mar1.lightcurve} shows
the resulting light curves, with the data taken from the Cambridge Photometry Calibration Server\footnote{\url{https://gsaweb.ast.cam.ac.uk/alerts/alert/Gaia17bpi}}
and the NEOWISE-R Single Exposure Source Table\footnote{ \url{ https://irsa.ipac.caltech.edu/}}
with the time
offset to provide a reasonable
match to the NEOWISE magnitudes. The overall
behavior is comparable to the observations, except that the
optical rise is much faster and sharper
than seen in the observations, and
the plateau phase is too faint by a little more than one magnitude.
Within the context of our models,
there is really no way of avoiding a
sharp optical peak; the onset of the
temperature perturbation
always results in a snowplow making
a very large, concentrated surface density
spike. Note that the immediate drop in the light curves is due to the initial conditions at the very inner edge of the disc which are not
consistent with steady state. These produce a slightly higher accretion rate at first, which dies out quickly.

We were able to increase the
optical flux with the same
structural model but doubling
the surface density of the
flat region, to
$\Sigma = 4000 \, \gmcmtwo$.
As shown in Figure 
\ref{fig:mar6lightcurve},
this results in much better
agreement with the optical
light curve, with a peak
temperature of $\sim 7300$~K,
but now
W1 and W2 are too bright by
about one magnitude. In general
we were unable to find
a model that substantially
changed the infrared/optical
flux ratio. 
As discussed in the Appendix,
there is a feature in the
model temperature distribution
that differs from the simple
$T \propto R^{-3/4}$ steady
disc solution, and this can
account for part of the discrepancy. Alternatively,
it may be possible that the
actual extinction is lower than
$A_V = 3.5$ 
\citep[see discussion in][]{hillenbrand18}, or
possibly that the IR-optical
extinction law is flatter than
the standard \citep{wang19}. Finally,
of course, there may be limitations of our simple treatment of time-dependent
disc accretion.

\begin{figure}
    \centering
    \includegraphics[width=0.45\textwidth]{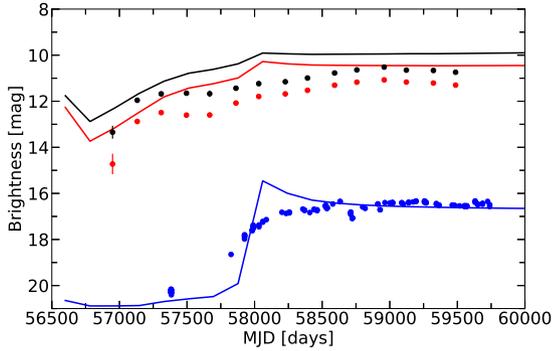}
    \caption{Light curves for a model with the same parameters
    as those in Figures \ref{fig:mar1structure}, \ref{fig:mar1_mdot},
    and \ref{fig:mar1.lightcurve}, except with twice the flat region
    surface density. The dots and curves are as in Figure \ref{fig:mar1.lightcurve}.
    }
    \label{fig:mar6lightcurve}
\end{figure}

\begin{figure*}
    \centering
    \includegraphics[width=0.95\textwidth]
    {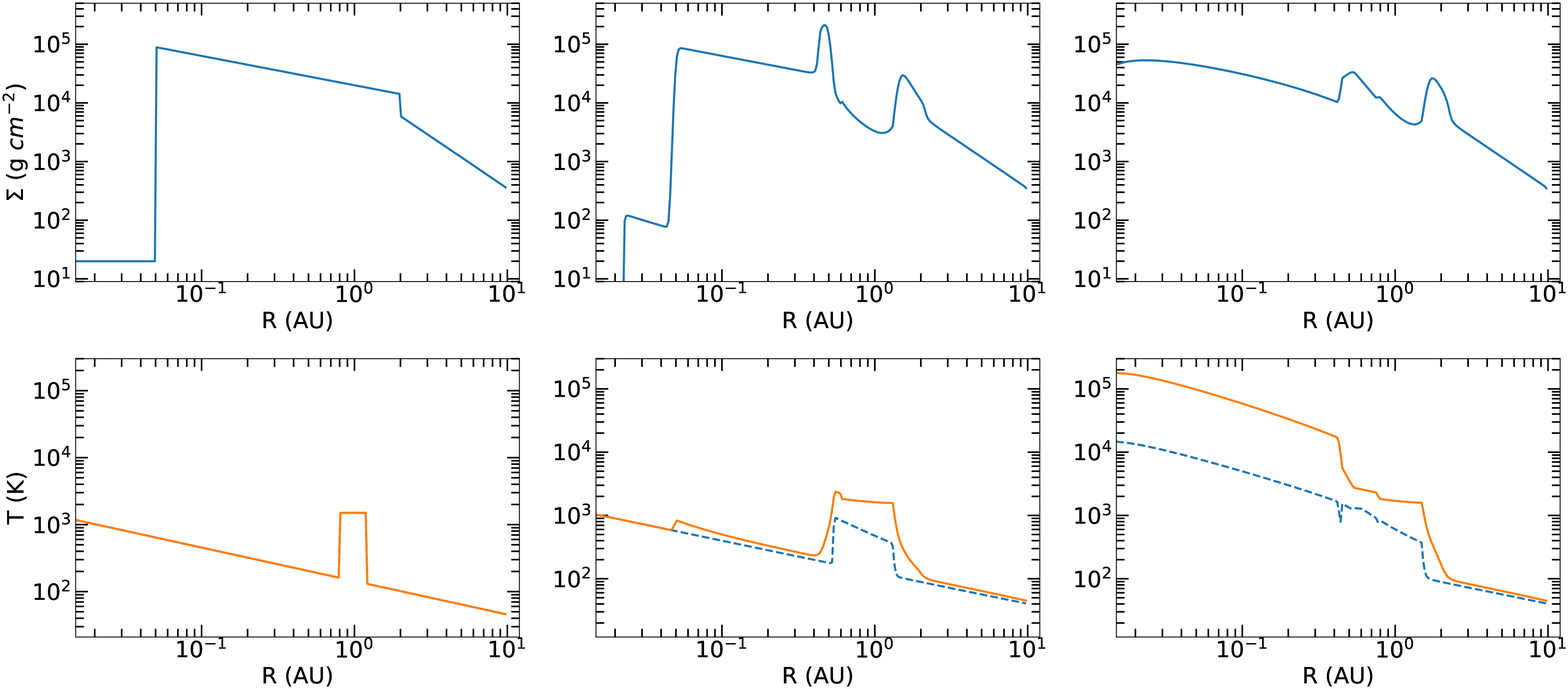}
    \caption{Evolution of an outburst model similar to FU Ori, showing the surface density structure and temperature at three times; the initial condition, the development of the outburst, and the structure at peak accretion rate. The dashed and solid curves show the surface (effective) temperature and the central temperature, respectively. From left to right, the structures are shown at 0, 70, and 95 years after the start of the model.}
    \label{fig:fuoristructure}
\end{figure*}

A brief exploration of other parameter choices is presented
in the Appendix. As shown there, the time lag between infrared and optical
outbursts is somewhat sensitive to the position of the temperature
perturbation.
Moreover,
values of the alpha parameter
that are three times smaller or larger lead to lag times between
the infrared and optical outbursts
that are much too large or too small
respectively. Overall, we conclude
that the propagation of the
outburst can be explained in
with $\alpha = 0.1$, in agreement with
the estimate of \cite{liu22}.

\begin{figure}
    \centering
    \includegraphics[width=0.45\textwidth]{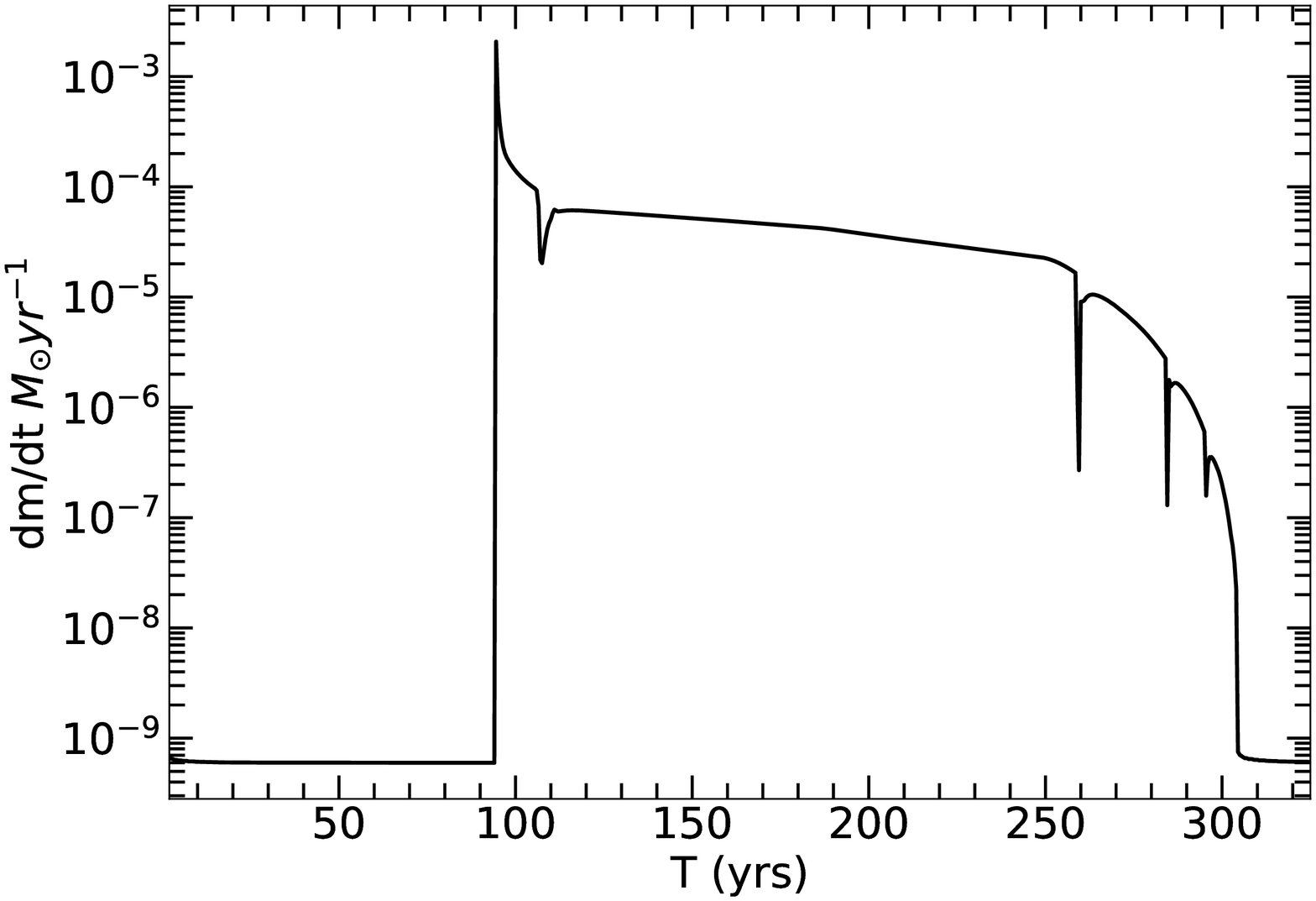}
    \caption{Accretion rate vs. time for the model in Figure \ref{fig:fuoristructure}.}
    \label{fig:fumdot}
\end{figure}

\subsection{Strong, FU Ori-like outburst}

\label{sec:fuori}

To simulate a large 
FU Ori-like outburst, we constructed 
a model with an initial condition similar 
to that in
in Figures 
\ref{fig:dec82_temps} 
and
\ref{fig:dec82_sigmas}
with the temperature perturbation
placed at 1 au.
This is similar to our previous models of FU Ori outbursts
resulting from continuing infall to the disc, 
which generates GI turbulent heating and subsequently
MRI activation at radii $\sim 3$~au \citep{zhu09,zhu10b,bae13,bae14}.
However, in this case the outburst is triggered by hand
rather than by GI. 

In contrast to the model shown in 
Figures 
\ref{fig:dec82_temps} and
\ref{fig:dec82_sigmas},
we decided to explore the effect of
a rising surface density $\Sigma \propto R^{-1/2}$ 
interior to the temperature perturbation (see Figure \ref{fig:fuoristructure}).
(A  steeper dependence on radius results in triggering
the MRI at very small radii, resulting in an inside-out
burst). 
We also impose an outer surface density 
distribution decreasing as $\Sigma \propto R^{-7/4}$.
This is done to avoid a gravitationally-unstable outer disc;
with $T \propto R^{-1/2}$, the Toomre parameter
$Q \propto T^{1/2} \Omega /\Sigma $ remains constant
(and sufficiently large).

The evolution of the temperature and surface
density is very similar to the previous model. 
The main difference is that 
the higher inner surface density produces accretion rates that
are about 2-3 times larger during the plateau phase, with higher accretion rates and a longer
outburst (compare Figure \ref{fig:fumdot} with
Figure \ref{fig:dec82_mdot}).

\begin{figure}
    \centering
    \includegraphics[width=0.45\textwidth]{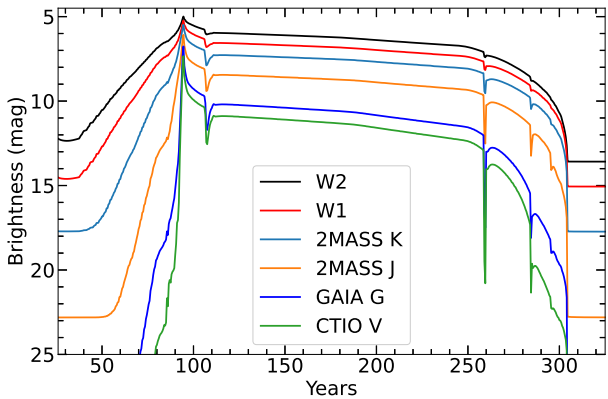}
    \caption{Light curves for a model FU Ori outburst. The light curves, from top to bottom, correspond in order to that shown in the legend.}
    \label{fig:fuorilightcurve}
\end{figure}

As in the case of Gaia 17bpi, the light curves (Figure \ref{fig:fuorilightcurve}) show that the infrared outburst
precedes the optical event, as the accretion wave moves
inward and the disc temperatures rise. However, in this case the
lag time is much larger, with the infrared rise starting about
40 years before the optical brightening. 
consistent with a longer timescale
for an outburst starting at much larger radii.
The mean propagation speed of the accretion wave is
$v_p \sim 1$~au$/40$~yr $\sim 0.1 \kms$. 
With a sound speed at 1500 K of $2.3 \kms$,
$v_p$ is of the order $\alpha c_s$, as expected from
$\alpha$-disc theory \citep[e.g.,][]{bell94}.
The actual propagation is not steady, but speeds up
with time as the disc temperature rises.

\section{Discussion}

\label{sec:discussion}

\subsection{Small outbursts: Gaia 17bpi}

Our models indicate that small
outbursts can be explained by MRI activation of low surface density
inner discs. These are very unlikely
to be produced by GI in discs, though
other mechanisms (see Introduction) remain. GI triggering at large radii results in high accretion rates.
These large outbursts, if MRI-driven,
are likely to be confined to disc radii within a few AU, as the required trapping of heat would
imply extremely large surface densities that would be so gravitationally-unstable that they would likely evolve on dynamical timescales.
Note that we
adopted the lowest surface density at each radius that sustains the MRI; larger surface densities would would produce larger
accretion rates.

In the case of Gaia 17bpi, we imposed
a low surface density in the innermost disc
prior to outburst, attributing that to
the usual magnetospheric truncation.
However, we cannot achieve the
necessary peak optical spectral type
or effective temperature (given the
accretion rate or luminosity) unless
the magnetosphere is crushed or otherwise absent, with an inner
radius close to the $\sim 0.65 \rsun$ estimated
by RH22. The absence of the 
usual broad optical emission
lines characteristic of 
magnetospheric
flows supports the absence of disc truncation.

Standard theory predicts that
the magnetospheric truncation radius 
is given by (e.g., K\"onigl 1991)
\begin{equation}
    R_m = \beta B^{4/7} R^{12/7}
    \mdot^{-2/7} (2 G M )^{-1/7}\,,
\end{equation}
where $B$ is the surface
(dipolar) field, $R$ and $M$ are the stellar radius and mass respectively,
and $\mdot$ is the mass accretion rate.
 The magnetosphere is
crushed when $R_m = R$, or
\begin{eqnarray}
    \mdot_c & = & \beta^{7/2} B^2 R^{5/2} (G M )^{-1/2}\\
    &= & 10^{-6} \beta_{0.5}^{7/2} B_{1.5}^2 R_{1.5}^{5/2} (G M_{0.5} )^{-1/2} \msunyr
    \,,
    \label{eq:magnetonumbers}
\end{eqnarray}
where the fiducial values are
$\beta_{0.5} = 0.5$, $B_{1.5} = 1500$~G, $R_{1.5} = 1.5\rsun$, and
$M_{0.5} = 0.5 \msun$.
The parameter $\beta$ is poorly known;
we take a fiducial value of 0.5 \citep{konigl89}, which with the other fiducial parameters yields a typically
estimated magnetospheric radius of
$\sim 5 R_*$ for a median T Tauri
accretion rate of $10^{-8} \msunyr$
and a stellar mass and radius of
$0.5 \msun$ and $2 \rsun$.
Thus, for the classical FU Ori objects, with accretion rates
$\gtrsim 10^{-5} \msunyr$, one would
expect the magnetosphere to be
suppressed, consistent with the observed lack of magnetopheric
emission lines.

At first, it seems surprising
that the low accretion rate implied for
Gaia 17bpi
can crush the magnetosphere.
However, RH22
estimated a remarkably small inner
disc radius for this object
$\sim 0.36 - 0.48 \rsun$. Assuming the
other fiducial parameters in
equation \ref{eq:magnetonumbers},
a stellar radius of $ 0.5 \rsun$ yields
$\mdot_c \sim 6 \times 10^{-8} \msunyr$.
In other words, a fixed magnetic field strength
on a much smaller star yields a much smaller
magnetic moment and less magnetospheric resistance to accretion.

\begin{figure}
    \centering
    \includegraphics[width=0.45\textwidth]
    {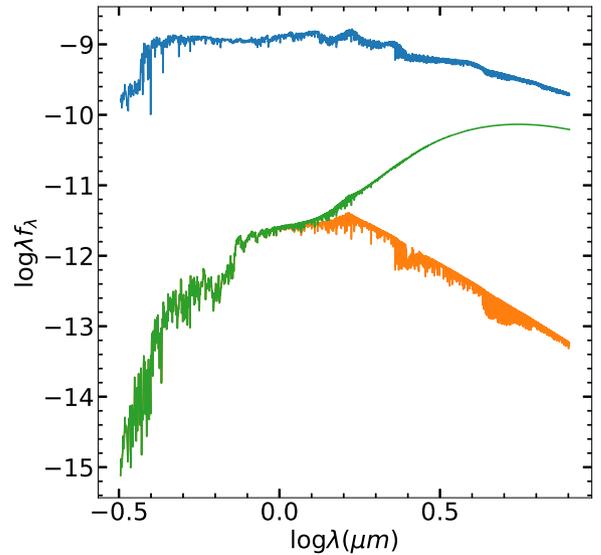}
    \caption{SEDs of FU Ori-type model, shown at
    $t = 70$ and 100 yr (see Figure \ref{fig:fuorilightcurve}). At 
    $t=70$ yr, the SED (green curve)
    is dominated by the stellar photosphere out to almost
    $\lambda = 2 \mu$m, but exhibits a large infrared excess compared to the stellar emission (orange spectrum. (The green curve overlaps the orange exactly at
    shorter wavelengths.)
     The upper (blue) curve shows the spectrum in full outburst at $t = 160$~yr, similar to that observed in FU Ori.}
    \label{fig:fu_sed_evol}
\end{figure}

\subsection{Large, FU Ori-like outbursts}

Our results are similar to those of
our previous models of GI triggering of the MRI
in discs on scales of order 1-3 au \citep{zhu10b,bae13,bae14}.
The simulations predict a much sharper optical rise in brightness
compared with observations.
This may be an artifact of our
assumption of axisymmetry and neglect of convection (see 
\cite{zhu10a}. Moreover, the
present treatment of MRI activation is of course
schematic; it may not capture the
real physics of the onset of magnetic
turbulent viscosity from an inert
state. 

The innermost disc effective temperatures
that we predict tend to be larger
than typically observed in outbursts
($\sim 10,000$~K vs. $\sim 6000$~K).
Again, the flow-through inner boundary adopted here means that
there is no turn-over in the
disc temperature distribution
at small radii characteristic of
the steady-disc solution, which
assumes a zero-torque boundary condition. We also have not taken
into account any possible energy loss via the ejection of winds,
which would reduce disc temperatures
\citep{zhu20}.
Peak temperatures are also sensitive to the inner disc radius for a given
accretion rate and stellar mass.
For example, in the case of 
FU Ori, SED modeling suggests that
the inner radius must be $\sim 3-4 \rsun$ \citep{hartmann96},
which is considerably larger
than the typical radii of $0.6 \msun$
pre-main sequence stars. It has been suggested that this is the result
of advection of large amounts of thermal energy into the central star, increasing its radius
\citep{baraffe12}. On the other hand, 
there are
outbursting objects whose optical spectra seem to be of higher
effective temperature and/or
weaker or absent absorption lines, or exhibit some composite signatures of ''hot''
and ''cool'' spectra \citep{hillenbrand19b,hillenbrand19a}.
High accretion rate 
objects with higher temperatures
also tend to have very strong
wind emission lines, which makes
them harder to classify and/or
interpret.

The most important result of our latest simulations
of large-scale outbursts is the finding that, assuming
$\alpha$ parameters similar to those estimated for
Gaia 17bpi, infrared outbursts can be detected many
years or even decades in advance of optical bursts, depending upon where the outburst
is triggered and the effective value of
$\alpha$.
This assumes outside-in propagation; V1515 Cyg shows a
slow optical rise that may suggest inside-out triggering,
in which case the optical and infrared rises will occur
roughly simultaneously \citep[e.g.,][]{bell95}.
In this context, the recent suggestion by \cite{nayakshin23}
that FU Ori outbursts are fueled by evaporation of hot Jupiters close to the central star would predict that the 
infrared light curve would not exhibit much lag behind the optical rise.
Multi-wavelength observations can thus provide a definitive test
of the two different scenarios.

Another possible test of the
outside-in large outburst picture would be the observation of large 
far-infrared excesses in objects that are in optical quiescence.
As shown in Figure \ref{fig:fu_sed_evol}, a large accretion rate burst originating at $\sim 1$~au would have a very large far-IR precursor. Distinguishing this behavior from other possibilities, such as radiation from a protostellar envelope, or an unresolved luminous companion, such as in
Z CMa \citep{koresko91,hinkley13}, can be difficult
but still may be worthy of consideration.

As noted in the Introduction, many theoretical simulations of FU Ori-type outbursts have been conducted, employing different triggering mechanisms.
It is difficult to consider the implications of most of
these studies for our work, because they generally assume inner radii far larger than considered here.
For example, the \cite{kadam20} study employs an
inner radius of 0.42 au, and most of the evolution and
emission that we calculate arises inside that boundary.
This is true even of our own previous work (for example, the inner radius is 0.2 au in \citealt{bae14}).

Our findings suggest that infrared surveys,
for example as might be provided by the {\em Roman Space 
Telescope} and other facilities, could provide
crucial insights into the physics of outbursts.
Whether
currently-detected variables
in the infrared 
\citep[e.g.][]{fischer19,fischer22}
will later have
optical bursts remains to be seen.
Further optical detections of bursts will obviously motivate
searches for prior infrared detections.

\section{Summary}

Using simple time-dependent models of disc outbursts, buttressed by
basic theoretical considerations, we find that
activation of the MRI can account for many
features of the diversity of observed accretion
events.
short-lived, relatively low accretion rate events
probably result from triggering in the inner disc,
and can occur at low surface densities, comparable
to or smaller than the minimum mass solar nebula,
and thus are very unlikely to result from MRI triggering
by gravitational instability. We find that the
time lag between the infrared precursor of the
outburst of Gaia 17bpi and the optical rise indicates
a rate of inward propagation consistent with 
$\alpha \sim 0.1$. We further find that larger outbursts, which must arise farther out in massive
discs, exhibit infrared precursors decades
before optical outbursts, raising the possibility
of detecting bursts early enough to study the
central star effectively, along with providing
insights into accretion physics.
These results emphasize the need for continued wide-field infrared monitoring, such as the VVV program
\citep{contreras14,lucas17,lucas20,park21,guo22} and NEOWISE \citep{park21,zakri22,contreras23} to complement 
the expected growth in optical outburst detections
over the next several years from all-sky surveys
such as from the {\em Vera Rubin Observatory}. 
Finally, three-dimensional non-
ideal magnetohydrodynamic simulations 
will be needed to take full advantage of
the observational results.

\section*{Acknowledgements}
We thank Lynne Hillenbrand for help with obtaining more recent light curves of Gaia 17bpi and an anonymous referee for a very helpful report.
This work was supported in part by the University of Michigan.
Our research has made use of the Spanish Virtual Observatory (\url{https://svo.cab.inta-csic.es)} project funded by MCIN/AEI/10.13039/501100011033/ through grant PID2020-112949GB-I00.
This work has made use of data from the European Space Agency (ESA)
mission {\it Gaia}
\citep{gaia16,gaia22}; 
(\url{https://www.cosmos.esa.int/gaia}) processed by
the {\it Gaia} Data Processing and Analysis Consortium (DPAC,
\url{https://www.cosmos.esa.int/web/gaia/dpac/consortium}). Funding
for the DPAC has been provided by national institutions, in particular
the institutions participating in the {\it Gaia} Multilateral Agreement.
This publication also makes use of data products from the Near-Earth Object Wide-field Infrared Survey Explorer (NEOWISE) \citep{mainzer14,cutri15}, which is a joint project of the Jet Propulsion Laboratory/California Institute of Technology and the University of Arizona. NEOWISE is funded by the National Aeronautics and Space Administration.

Software: PyAstronomy \citep{pya}, Spectres \citep{Spectres}, and Matplotlib \citep{Matplotlib}.

\section*{Data availability}

The models and associated data will be provided upon
reasonable request to the authors.

\bibliographystyle{mnras}
\bibliography{refs1}

\begin{figure*}
    \centering
    \includegraphics[width=1\textwidth]{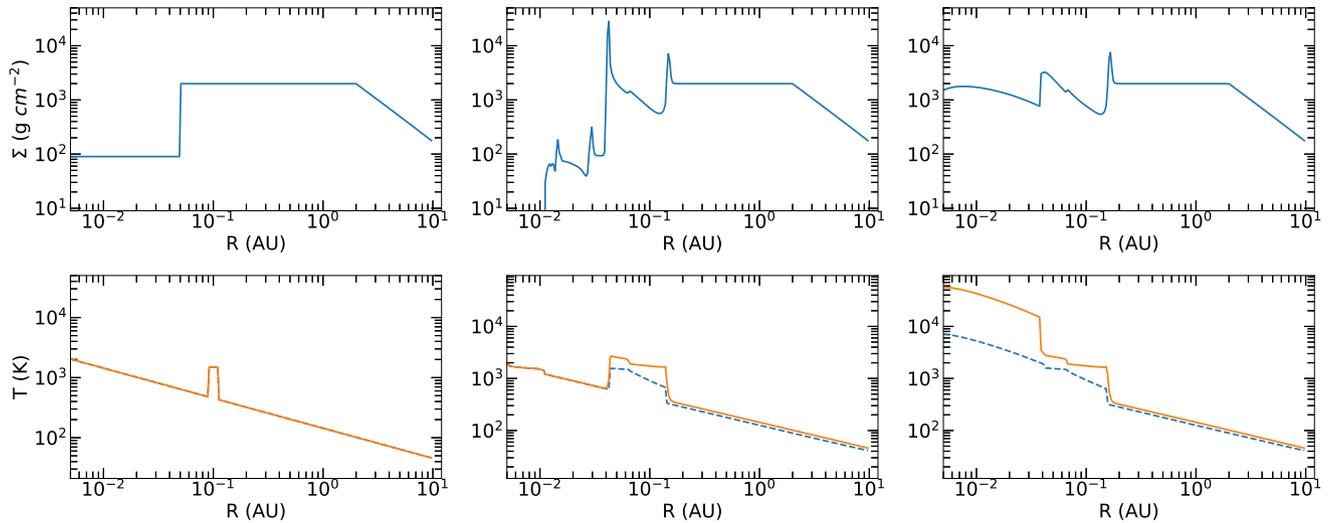}
    \caption{Initial conditions for a first attempt at modeling Gaia 17bpi (see text).}
    \label{fig:oct13structure}
\end{figure*}

\section*{Appendix}

Our first attempt at a model for Gaia 17bpi, assuming 
with $\alpha = 0.1$.
is shown in Figure \ref{fig:oct13structure}.
The inner disc radius is set at
0.005 au ($\sim 1 \rsun$). The surface density
is a constant $2000\, \gmcmtwo$ between 0.05 and 2 au, but with
a strong depletion inside of 0.05 au.  This is
to avoid having the innermost regions go into outburst
first, resulting in an outside-in burst
\S \ref{sec:methods}. We interpret our evacuated
inner region as corresponding to magnetospheric truncation.
The falloff of $\Sigma$ at large radii is intended to prevent
GI, but in any event the outburst cycle is not affected by
this structure.
The imposed temperature perturbation is centered at 0.10 au.
In summary, this initial structure is very similar to that
used for our ``final'' Gaia 17bpi model (Figure \ref{fig:mar1structure}. except the high
surface density region extends to larger radii.
The resulting evolution in Figures \ref{fig:oct13structure},
\ref{fig:oct13mdot}, and \ref{fig:oct13lightcurve} is very
similar to that shown in 
Figures \ref{fig:mar1structure}, \ref{fig:mar1_mdot}, and \ref{fig:mar1.lightcurve}.

\begin{figure}
    \centering
    \includegraphics[width=0.40\textwidth]{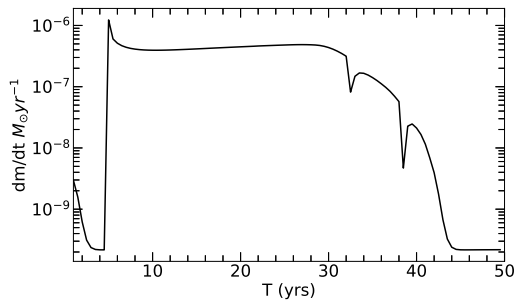}
    \caption{Accretion rate vs. time for model shown in Figure \ref{fig:oct13structure}. }
    \label{fig:oct13mdot}
\end{figure}

\begin{figure}
    \centering
    \includegraphics[width=0.45\textwidth]{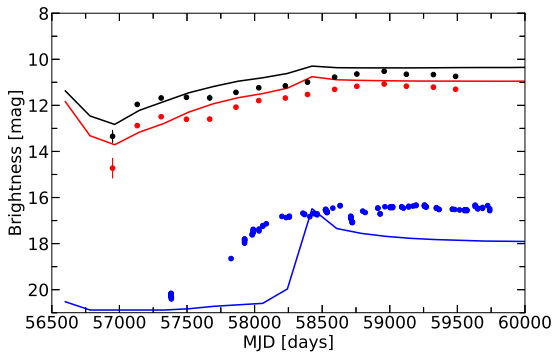}
    \caption{Light curves for the simulation shown
    in Figures \ref{fig:oct13structure} and \ref{fig:oct13mdot}.}
    \label{fig:oct13lightcurve}
\end{figure}

\begin{figure}
\includegraphics[width=0.45\textwidth]{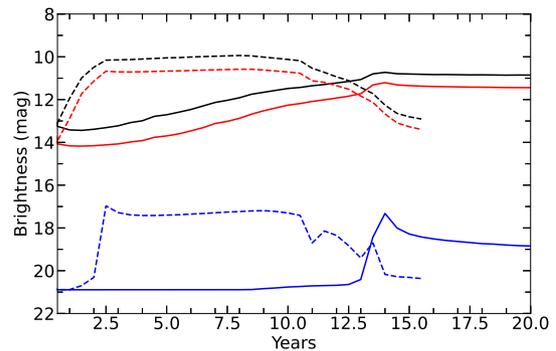}
\caption{Light curves
calculated for the same structural model $\Sigma$ and $T$ as in Figures \ref{fig:oct13structure} for
a viscosity parameter of $\alpha = 0.03$ (solid curves)
and a reduction in the surface density by a factor of three
for the $\alpha = 0.3$ case (dashed lines). 
The high viscosity results in a very
quick onset of the optically-visible burst, with very little
lag time from the infrared, while $\alpha = 0.03$
yields a much longer lag time than observed in Gaia 17bpi.}
\label{fig:alphatestlc}
\end{figure}

Figure \ref{fig:oct13lightcurve} shows a slightly larger
lag time between infrared and optical bursts than observed.
The difference with the results shown in Figure \ref{fig:mar1.lightcurve} is that in the latter case the
temperature perturbation was centered at 0.09 au instead of
0.10 au. However, the lag time is much more sensitive to
the adopted alpha parameter.
In Figure \ref{fig:alphatestlc} we show the effects of changing
the $\alpha$ parameter on the lag time between infrared and 
and optical light curves. (We reduced the surface density
for the large viscosity case because we wanted to keep a similar
accretion rate, and $\mdot \propto \alpha \Sigma$.) It is clear that $\alpha = 0.3$ yields
very little lag time between the infrared and optical light curves,
while $\alpha= 0.03$ results in a much longer lag than observed.

\begin{figure}
    \centering
    \includegraphics[width=0.45\textwidth]{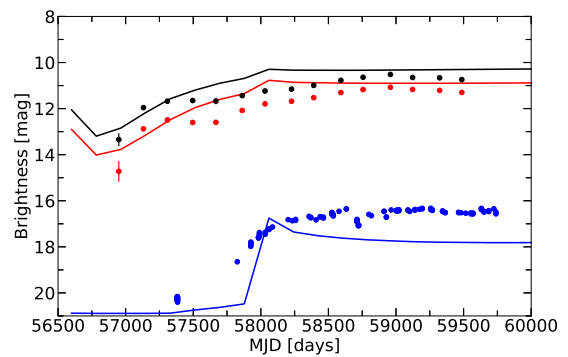}
    \caption{Light curve comparison
    with the same surface density distribution and temperature perturbation as in Figure 
    \ref{fig:oct13structure}, but lowering the temperature onset of the MRI to occur
    between 1000 K and 1100 K (see text).}
    \label{fig:gaialowt}
\end{figure}

Another issue is that the WISE fluxes are too bright relative
while the G fluxes are low. 
We lowered the activation temperatures
from 1200 K - 1300 K to 1000 K - 1100 K to see if this
would reduce the brightness in the WISE
bands while leaving the Gaia light curve
unchanged. Figure \ref{fig:gaialowt} shows
that the W1 and W2 magnitudes became only slightly fainter.
The main effect was to have the optical outburst occur sooner,
in better agreement with observations. 

Finally, we used the same model parameters but
setting the
flat surface density region at $\Sigma = 4000 \, \gmcmtwo$ rather than at $2000 \, \gmcmtwo$.
As expected, this results in a higher accretion rate
by roughly a factor of two (Figure \ref{fig:mar6lightcurve}).
The G magnitudes 
are in better agreement
with observations, but now W1 and W2 are too bright.


These results show that we have been unable to significantly
change the ratio of disc infrared and optical fluxes
during outburst. Changes in accretion rate simply
end up being reflected in the emission at both wavelength
regions. This isn't particularly surprising to the extent that the disc temperature distribution during outburst
is similar to that of a steady disc. However, there is a subtlety in the model calculations that enhances the
infrared relative to the optical in comparison with a steady disc. As shown in Figure \ref{fig:mar1.texcess},
the time-dependent models show an elevated surface
temperature relative to a steady-disc power-law distribution.
The reason for this is that the adopted dust opacity
disappears above $T \sim 1500$~K. Since the gas opacity
at slightly higher temperatures is quite small, this
results in pinning the temperatures near this transition.
This would not matter if the accretion rate 
was precisely independent of radius,
but it is time-dependent. This region of elevated
temperature results in the WISE fluxes being about
0.5 mag brighter than would be the case for the dot-dash
temperature distribution. 

\begin{figure}
    \centering
    \includegraphics[width=0.45\textwidth]{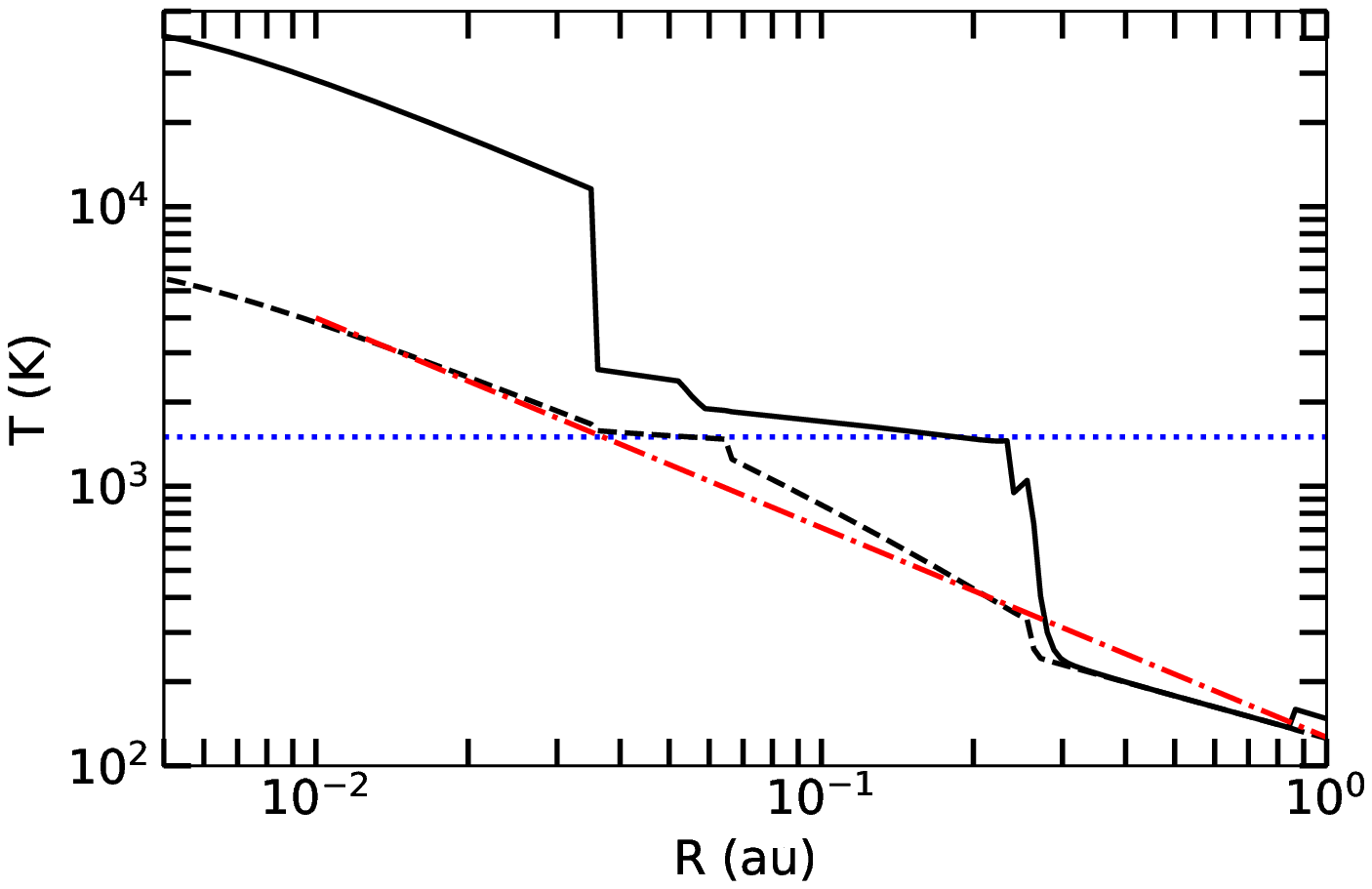}
    \caption{Comparison of disc temperatures for the
    model in Figures 
        \ref{fig:mar1structure} 
        and \ref{fig:mar1.lightcurve}
    with that of a simple steady
    disc $T \propto R^{-3/4}$ at a time
    $t =20$~yr. The model exhibits an
    excess relative to the power-law behavior between about 
    0.4 and 2 au (see text).}
    \label{fig:mar1.texcess}
\end{figure}

In summary, our results show that
the behavior of Gaia 17bpi can
be reasonably approximated by a temperature perturbation activating
the MRI at a radius $\sim 0.1$~au and
$\alpha = 0.1$. These results are similar to the estimate of \cite{liu22} of the same alpha
and a starting radius of 0.07 au
based on a simple model of
outburst propagation. The ratio
of infrared to optical emission
in our models is larger than
observed if $A_V = 3.5$; decreasing
this by $\sim$ 1 magnitude would
result in much better agreement,
and might be within observational
uncertainties \cite[see, e.g.][]{hillenbrand18} The discrepancy
might also reflect limitations
in our 1-dimensional alpha models,
which predict a sharper optical
rise than observed.

\label{lastpage}

\end{document}